# Fluoroscopy-based Navigation System in Orthopaedic Surgery

MERLOZ Ph (1), TROCCAZ J (2), VOUAILLAT H (1), VASILE Ch (1), TONETTI J (1), EID A (1), PLAWESKI S (1).

(1) University Department of Orthopaedic and Trauma Surgery ; CHU A. Michallon ; BP 217, 38043 Grenoble cedex 09 (France)

Tel : + 33 4 76 76 55 93 ; Fax : + 33 4 76 76 52 18 ; Pmerloz@chu-grenoble.fr

(2) Equipe GMCAO - Laboratoire TIMC/IMAG (Université Joseph Fourier - CNRS UMR 5525). Institut d'Ingénierie de l'Information de Santé
Faculté de Médecine - 38706 La Tronche cedex - France
Tel: +33 (0)4 56 52 00 06 - Fax: +33 (0)4 56 52 00 55
Secr: +33 (0)4 56 52 00 07 ; Jocelyne.troccaz@imag.fr

http://www-timc.imag.fr/gmcao et http://www-timc.imag.fr/Jocelyne.Troccaz

**Summary :**

This paper describes a computer-assisted surgical navigation system based on fluoroscopic X-ray image calibration and 3D optical localizers in order to reduce radiation exposure while increasing accuracy and reliablility of the surgical procedure for implant insertion.

The system allows real-time navigation in several X-ray projections simultaneously, with the fluoroscope turned off and removed from the operating field. In Trauma surgery procedures such as osteosynthesis of Ilio sacral joint, pedicle screw placement in spine, distal locking of intramedullary nails or, extra – intra capsular femoral neck fractures fixation, safety and accuracy can be improved thanks to the multiplanar guidance, while radiation exposure of both patient and surgical staff can be significantly reduced.

The virtual fluoroscopy show clearly that the system provide better accuracy and safety than conventionnal fluoroscopy. When we compared irradiation data with the same data collected during conventional procedures we can say that virtual fluoroscopy reduces significantly radiation exposure by the 1 / 3 ratio. The operative time with virtual fluoroscopy is slightly greater than the one recorded with conventional techniques. A fluoroscopy-based computer system can be seen as a complement to a CT-based computer system (CT scans provide full 3D image data, not fluoroscopic images). The advantages over CT-based systems are twofold : instant availability without preoperative preparation (no CT acquisition required) and up-to-date image data of patient anatomy (X-ray images used for navigation are acquired at the beginning of surgical procedure). As compared to standard fluoroscopy, the fluoroscopy-based computer system allows real-time navigation in several X-ray projections simultaneously and reduces significantly radiation exposure. Finally, the authors believe that this system can also greatly improve on surgical accuracy and safety of many applications in orthopaedics.

I. Introduction :

Mobile fluoroscopic device is an integral part of the standard equipment used in orthopaedic surgery to provide real-time feedback of bone and surgical tool positions. One of the disadvantages of this technique include the need for continuous radiation exposure for real-time visual control. For instance, in some specific surgical procedures such as distal locking in standard closed intra-medullary nailing of the femur and tibia, it is well known that this procedure is time-consuming and causes much irradiation to the patient and to the surgical team. In this exemple the surgeon's direct exposure to radiation varies from 3.1 to 31.4 min per procedure and 31 to 51 % of the total irradiation is spent on distal locking alone [1, 2, 3]. In clinical practice, mobile fluoroscopy device is not only used for distal locking in standard closed intra-medullary nailing for femur and/or tibial fractures, but also for many others surgical procedures. For instance, three other classical procedures are well known to cause much radiation : insertion of pedicle screws in spine surgery for trauma, scoliosis and degenerative instabilities ; osteosynthesis of intra - extra capsular femoral neck fractures and ilio-sacral joint fixation for the treatment of ilio-sacral joint disruption. This paper describes a computer-assisted surgical navigation system based on fluoroscopic X-ray image calibration and 3D optical localizers in order to reduce radiation exposure while increasing accuracy and reliablility of the surgical procedure for implant insertion. We will describe the surgical technique used for pedicle screw placement in spine surgery, before a brief report on other techniques.

II. Material and Methods :

A three-dimensional optical localizer (*Polaris system ; Northern Digital ; Ontario ; Canada*) is used to track the position of surgical tools, patient reference and C-arm image intensifier (*OEC Medical system ; Courtabeuf ; France*) within the region of the operating table (Fig.1). Each component is equipped with passive markers that give the position of three distinct tools (patient reference frame ; surgical tools : *pointer, sharp tip probe, bradawl and drill guide* ; C-arm), with the help of the three - dimensional optical localizer.

Fluoroscopic images are usually distorted. This is due to three factors: the planarity of the image intensifier input screen, the insufficient rigidity of the c-arm and the surrounding magnetic field [4, 5]. To correct image intensifier distortions and calibrate images, a calibration grid (equipped with passive markers) is attached rigidly to the input screen of the image intensifier (Fig.2). The aim of the calibration process is to learn the mapping between image pixels and the "physical" surgical space, so that the computer can generate a virtual projection of the surgeon's tool axis onto the calibrated X-ray views. The calibration device is fixed to the image intensifier and contains radio-opaque elements arranged on two parallel plates. Radio-opaque elements who are included in the lower plate are used to correct the image distortions and these of the upper one to detemine the x-rays source position (Fig.2). To determine the x-ray source position, passive markers are fixed on the calibration device and the three-dimensional optical tracking system is used to localize the fluoroscopic imaging device (Fig.2, 3).

The correction of original images can be effected thanks to the radio-opaque elements of the lower plate projected onto the images (Fig.4).

We describe here the surgical technique using virtual fluoroscopy in spine surgery for pedicle screw fixation. In spine surgery, the complications associated with misplaced pedicle screws are mostly neurological or vascular. Previous study of surgical procedures have shown a significant rate of incorrect placement of pedicle screws ranging from 15 % to 40 % [6, 7]. Virtual fluoroscopy is one of the solution to increase safety of screw placement.

The patient is settled in prone position and the surgeon perform a posterior approach. In a first step, the surgeon attaches the dynamic reference frame to the patient's bone of interest (in this case : spinous process of the vertebra to be operated on) and aligns the optical localizer's cameras optimally (Fig.3). In a second step, the surgeon acquires one X-ray view, just to calibrate the C arm. In a third step, two single X-ray views from A-P and lateral positions are acquired within the area who has to be operated on (Fig.5). Data are passed on to the workstation computer (*Fluologics system, Praxim, France*) (Fig.1). After calibration of the acquired X-rays by the computer, the fluoroscope is turned off and removed from the operating field.

Finally, the calibrated views are displayed on the workstation screen. A computer-generated projection of the surgeon's surgical tool is also displayed in each view. This is equivalent to its representation under conventional constant fluoroscopic control. A real-time navigation in several views is possible simultaneously (Fig.6). The feasibility of this technique for lumbar pedicle screw insertion was assessed through cadaver and clinical trials by Foley et al. [7]. Twenty six patients underwent posterior osteosynthesis with pedicle screws to correct fractures and degenerative instability of the thoraco-lumbar spine by using virtual fluoroscopy (26 patients ; vertebra level: T10 to L5 ; 140 screws). In the same time, twenty six patients underwent the same surgical procedure (26 patients ; vertebra level: T10 to L5 ; 124 screws) by using conventional technique (surgical classical procedure to insert pedicle screws, and immediate fluoroscopic control). Evaluation of screw placement in every cases was done by using plain X rays and post operative CT scan, with the help of an independant examiner. Irradiation data and intervention duration were collected during conventional and computer-aided procedures.

III.     Results :

In the computer-aided group (26 patients ; vertebra level: T10 to L5 ; 140 pedicle screws) the results showed that the technique provided much better safety and accuracy than manual insertion (Fig.7). A percentage of cortex penetration of less 5 % (7 of 140 pedicle screws) occured for the computer assisted group with an average below or equal to 2 mm in five cases. In the two other cases the cortex penetration was slightly higher than 2 mm without complications. A percentage of penetration of 13 % (18 of 124 pedicle screws) occured for the non computer assisted group (26 patients ; vertebra level: T10 to L5 ; 124 pedicle screws) with an average below or equal to 2 mm in four cases, and with an average below or equal to 3 mm in 14 cases . In the two groups there were no neurologic complications.

The virtual fluoroscopy shows clearly that the system provides better accuracy and safety than conventional fluoroscopy. When we compare irradiation data with the same data collected during conventional procedures we can say that virtual fluoroscopy reduces significantly radiation exposure time by the 1 / 3 ratio (3,5 sec. on the average for each procedure with virtual fluoroscopy *versus* 11,5 sec. on the average for each procedure with conventional technique). The operative time with virtual fluoroscopy is slightly greater than the one recorded with conventional techniques (10 minutes on the average for one vertebra level - two screws - with conventional procedure *versus* 11,9 minutes on the average for one vertebra level - two screws - with virtual fluoroscopy). The main reasons for time increase are the following : setup, reference frame attachment, data acquisition, significant learning curve.

IV.    Discussion :

This paper describes a computer-assisted surgical navigation system based on fluoroscopic X-ray image calibration and 3D optical localizers. Currently available computer-assisted orthopaedic systems are generally based on 3D image data sets that are acquired preoperatively with a Computed Tomography (CT) scanner [6, 8]. A fluoroscopy-based computer system can be seen as a complement or an alternative to the use of a CT-based computer system (CT scans provide full 3D image data, not fluoroscopic images). The advantages over CT-based systems are twofold : instant availability without preoperative preparation (no CT acquisition required) and up-to-date image data of patient anatomy (X-ray images used for navigation are acquired at the beginning of surgical procedure) [6, 7, 8]. As compared to standard fluoroscopy, the fluoroscopy-based computer system allows real-time navigation in several X-ray projections simultaneously, with the fluoroscope turned off and removed from the operating field and reduces significantly radiation exposure of both patient and surgical staff. Finally, we believe that this system can also greatly improve on surgical accuracy and safety of other applications in orthopaedics [9, 10].

Fluoroscopy is commonly used in some other surgical procedures such as distal locking in closed intra-medullary nailing for femur and/or tibial fractures, osteosynthesis of intra - extra capsular femoral neck fractures and ilio-sacral joint fixation for pelvic ring disruption. With conventional procedure, each of them represents a real clinical challenge with technical complexity and potentially leading to additional pitfalls and complications and necessitates some specific technical details.

For instance, percutaneous distal locking is a good challenging step in standard closed intra-medullary nailing of the femur. Many complications can be described with distal locking procedure : inadequate fixation, malrotation, bone cracking, cortical wall penetration, bone weakening due to multiple or enlarged screw holes and distal fragment rotation if the drill is not perfectly aligned with the hole axis in the nail. Improving screw placements, virtual fluoroscopy leads to reduce the complication rate [10, 11]. Taking into account technical aspects, when using virtual fluoroscopy for distal locking in closed intra-medullary nailing of the femur the reference frame has to be fixed onto the distal part of the femur (metaphysis).

Ilio-sacral joint screwing is another difficult clinical challenge, and some intra and postoperative complications can occured, such as inadequate fixation or cortical wall penetration [9, 12]. If it's so, an extra osseous course in the anterior part of the sacrum, can lead to injure the lumbo-sacral root of the ischiatic nerve. Safety and accuracy can be improved thanks to the "virtual" multiplanar guidance, while radiation exposure can be also significantly reduced [13, 14, 15, 16]. To optimize ilio-sacral joint screwing when using virtual fluoroscopy the reference frame is generally attached to the iliac crest, and the procedure in most of the cases can be performed percutaneously.

Inadequate fixation and cortical wall penetration are the two main complications when performing osteosynthesis of intra – extra capsular femoral neck fractures. Once again, the use of virtual fluoroscopy demonstrated that this technique could increase safety and accuracy in osteosynthesis and the procedure is generally performed with a minimal invasive approach. In this example (trochanteric and femoral neck fractures), the reference frame has to be fixed onto the proximal part of the femur (trochanteric area).

Regarding the basic principle of CAS systems, i.e. multi-modal data registration, the authors wish to focus the attention on some new technologies that are currently under development or being tested, which they believe will soon replace the existing techniques in the next generation of CAS systems, mainly because of their increased accuracy and efficiency, but also because of an easier implementation in the OR [17]. The recent coming out of 3D fluoroscopic devices [18, 19, 20] should allow surgeons to dispose of 3D models without the need for any data registration at all. This new technology should allow percutaneous implant (screw) placement to be performed in a very near future [21].

V.     Conclusion :

A reduction in outliers (screw misplacement) was consistently associated with the use of computer-assisted techniques, regardless of the navigation system used. Many studies indicate that the consistency of screw placement achieved with computer-assisted techniques results in a corresponding consistency in clinical functional outcomes. This paper emphasizes the importance of examining the incidence of outlier functional results in the evaluation of computer-assisted surgery. Cost, additional time, and learning curve might very well be compensated by the significant outlier reduction.

**REFERENCES**

[1] Coetzee J.C. and van der Merwe E.J., "Exposure of surgeons-in-training to radiation during intramedullary fixation of femoral shaft fractures," S Afr Med J, 1992. 81(6): p. 312-4.

[2] Levin P.E., Schoen R.W. Jr, Browner B.D., "Radiation exposure to the surgeon during closed interlocking intramedullary nailing," J Bone Joint Surg Am, 1987. 69(5): p. 761-6.

[3] Skejdal S. and Backe S., "Interlocking medullary nails – radiation doses in distal targeting," Arch Orthop Trauma Surg, 1987. 106: p. 179-181.

[4] Chakraborty D.P., "Image Intensifier Distortion Correction," Medical Physics, 1987. 14(2): p. 249-252.

[5] Hoffmann K.R., Chen Y., Esthappan J., Chen S.Y.J., Carroll J.D., "Pincushion correction techniques and their effects on calculated 3D positions and imaging geometries," SPIE Medical Imaging: Image Processing, 1996. 2710: p. 462-467.

[6] Merloz Ph, Tonetti J, Eid A, Faure C, Lavallée S, Troccaz J, Sautot P, Hamadeh A, Cinquin Ph.

Computer Assisted Spine Surgery

Clin. Orthop. 1997, 337, 86-96.

[7] K.T. Foley KT, Simon DA, Rampersaud YR.

Virtual fluoroscopy.

Operative Techniques in Orthopaedics, 10, 1, 2000: 77-81

[8] Merloz Ph, Huberson Ch, Tonetti J, Eid A, Vouaillat H.

Computer-Assisted Pedicle Screw Insertion.

Techniques in Orthopaedics, Vol. 18 (2) : 149 – 159 ; 2003

Lippincott Williams & Wilkins, Inc., Philadephia

[9] Kahler DM, Mallik K.

Computer-Assisted Iliosacral screw placement of compared to standard fluoroscopic technique.

Comput Aided Surg 1999, 6, 348.

[10] Kahler DM.

Virtual fluoroscopy : a tool for decreasing radiation exposure during femoral intramedullary nailing.

Stud Health Techno Inform. 2001 ; 81 : 225-228

[11] Hazan EJ, Joskowicz L.

Computer-assisted image-guided intramedullary nailing of femoral shaft fractures.

Techniques in Orthopaedics, Vol. 18 (2) : 191 – 200 ; 2003

Lippincott Williams & Wilkins, Inc., Philadephia

[12] Lange R.H, Hansen ST.

Pelvic ring disruptions with symphysis pubis diastasis : indications, technique and limitations in anterior internal fixation. Clin. Orthop; 1985: 201, 130-137.

[13] Gautier E, Bachler R, Heini PF, Nolte LP.

Accuracy of computer-guided screw fixation of the sacroiliac joint.

Clin. Orthop 2001 ; 393: 310 – 317

[14] Schep NWL, Van den Bosch EW, Haverlag R, Van Vugt AB.

Positioning of sacroiliac screws with fluoroscopy based navigation technique. Proceedings of the second meeting of the international society for computer assisted orthopaedic surgery.

June 19-22, 2002, Santa Fé, New Mexico, USA, Digioia A, Nolte LP Eds., Carnegie mallon University, Pittsburgh, 2002, pp 312.

[15] Stöckle U, König B, Hofstetter R, Nolte LP, Hass NP.

Navigation assisted by image conversion. An experimental study on pelvic screw fixation.

Unfallchirurgg. 2001, 104 (4), 215-220.

[16] Grutzner PA, Rose E, Vock B, Holz F, Nolte LP, Wentzensen A.

Virtual fluoroscopy in accute treatment of pelvic ring disruptions

Unfallchirurg. 2002, 105, 254-260.

[17] J. Tonetti, L. Carrat, S. Blendea, Ph. Merloz, J. Troccaz, S. Lavallée, JP. Chirossel.

Clinical results of percutaneous pelvic surgery. Computer assisted surgery using ultrasound compared to standard fluoroscopy. Computer Aided Surgery, 2001, 6, 204-211

[18] Desbat L, Fleute M, Defrise M, Liu X, Huberson C, Laouar R, Martin R, Guillou JH, Lavallée S.

Minimally Invasive Interventional Imaging for Computer-Assisted Orthopaedic Surgery.

In Troccaz J, Merloz Ph, editors. "SURGETICA 2002", Computer-aided medical interventions: tools and applications, Montpellier : Sauramps Medical; 2002, p. 288 – 295

[19] Grutzner PA, Wälti H, Nolte LP.

First clinical experience in inherent registration techniques with mobile 3D fluoroscopy for matching free navigation. Preliminary report of a new method for intraoperative navigation. Proceedings of the second meeting of the international society for computer assisted orthopaedic surgery ; June 19-22, 2002, Santa Fé, New Mexico, USA, Digioia A, Nolte LP Eds., Carnegie mellon University, Pittsburgh, 2002, pp 220-221.

[20] Ritter D, Mitschke M

Direct Marker-free 3D Navigation with an Isocentric Mobile C-Arm.

In Troccaz J, Merloz Ph, editors. "SURGETICA 2002", Computer-aided medical interventions: tools and applications, Montpellier : Sauramps Medical; 2002, p. 288 – 295.

[21] Tonetti J, Carrat L, Blendea S, Merloz P, Troccaz J, Lavallee S, Chirossel JP

Clinical results of percutaneous pelvic surgery. Computer assisted surgery using ultrasound compared to standard fluoroscopy.

Comput. Aided Surg. 2001; 6 (4) : 204-211

**Figures and Legends :**

Fig. 1 : A three-dimensional optical localizer (A) is used to track the position of surgical tools, patient reference and C-arm image intensifier (B) within the region of the operating table. One can see the computer under the 3D localizer.

Fig. 2 : The calibration grid is fixed to the image intensifier and contains radio-opaque elements arranged on two parallel plates. Radio-opaque elements who are included in one of the plate (picture) are used to correct the image distortions.

Fig. 3 : To determine the x-ray source position, passive markers are fixed on the calibration device and the three-dimensional optical tracking system is used to localize the fluoroscopic imaging system. One can see the reference frame fixed onto the spinous process of the vertebra to be operated on ( ----------- ),  and the markers on the calibration grid.

Fig. 4 : The correction of original images (on the left) can be effected thanks to the radio-opaque elements of the lower plate projected onto the images (on the right).

Fig. 5 : Two single X-ray views from A-P and lateral positions are acquired within the area who has to be operated on (here a lumbar vertebra).

Fig. 6 : A computer-generated projection of the surgeon's surgical tool is displayed in each view. A real-time navigation in several views is possible simultaneously.

Fig. 7 : Evaluation of pedicle screw placement. L1 lumbar fracture (screws on T12, L1 and L2). A : conventionnal AP view ; B : conventionnal lateral view ; C : CT scan lateral view ; D : CT scan axial view on  T12). The two screws on T12 are perfectly inserted in the pedicle.

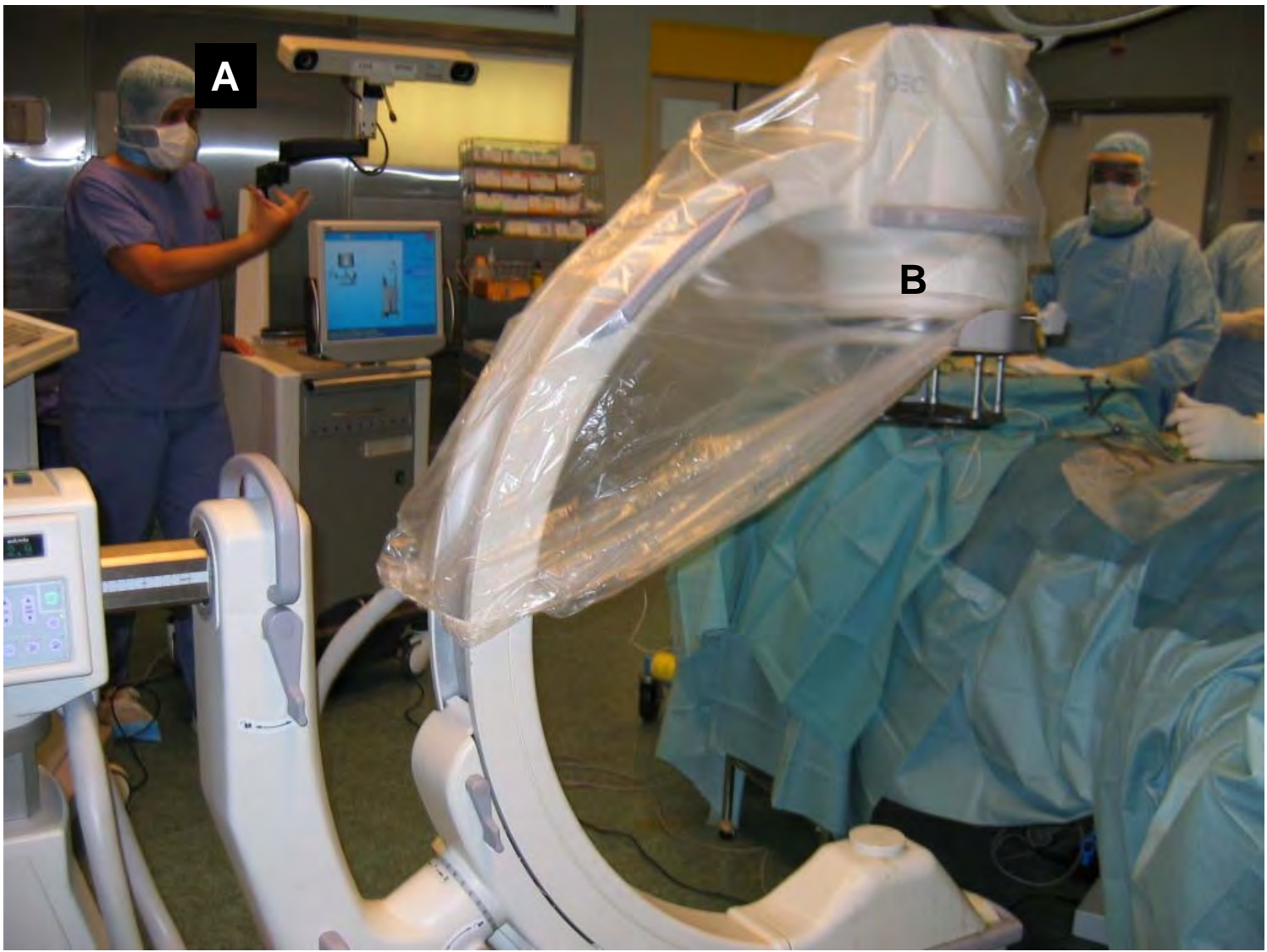

Fig. 1 : A three-dimensional optical localizer (A) is used to track the position of surgical tools, patient reference and C-arm image intensifier (B) within the region of the operating table.

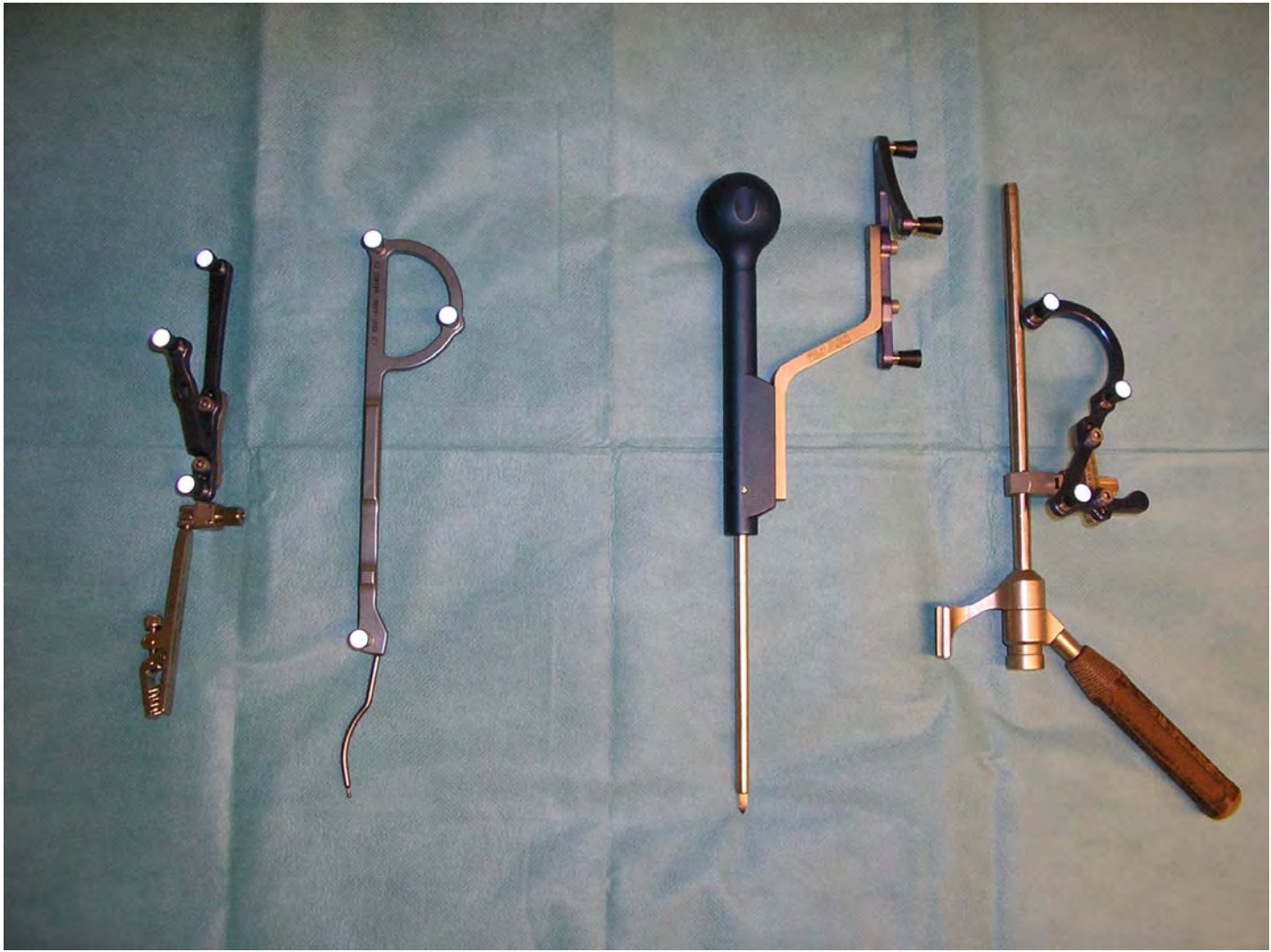

Fig. 2 : Navigated tools and the markers. Each tool is equipped with passive markers (white spots can be seen). From the left to the right, one can see the reference frame (fixed onto the spinous process of the vertebra to be operated on), the pointer, the bradawl and the drill guide.

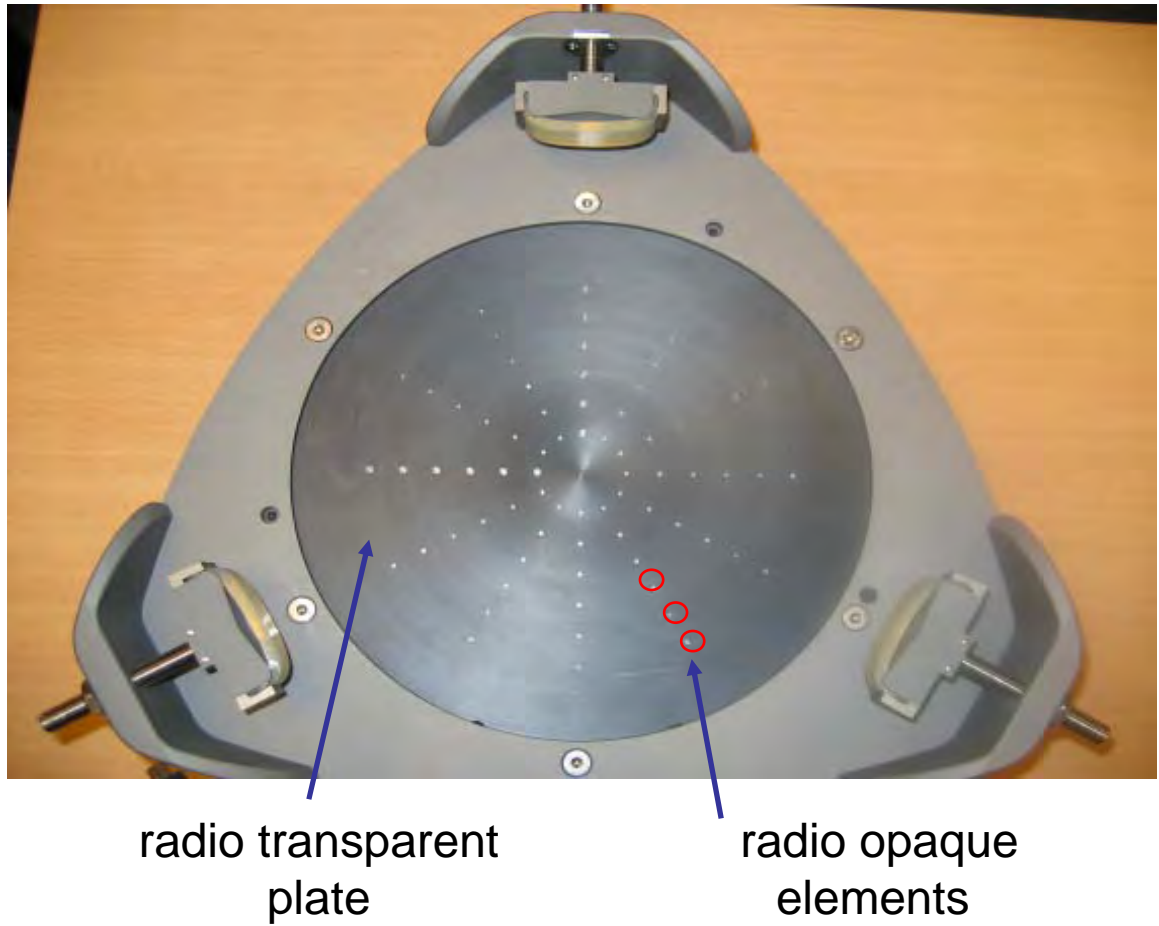

radio transparent plate    radio opaque elements

Fig. 3 : The calibration grid is fixed to the image intensifier and contains radio-opaque elements arranged on two parallel plates. Radio-opaque elements who are included in one of the plate (picture) are used to correct the image distortions.

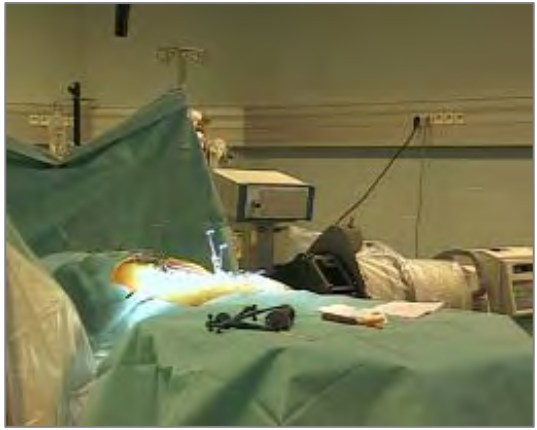
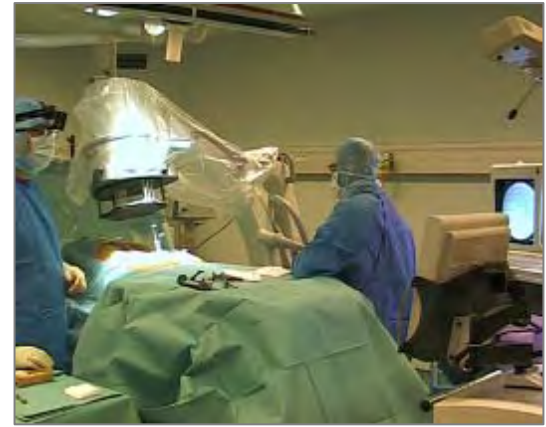
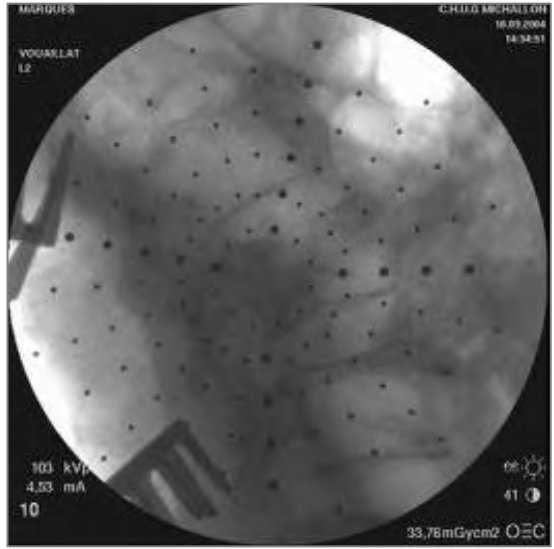
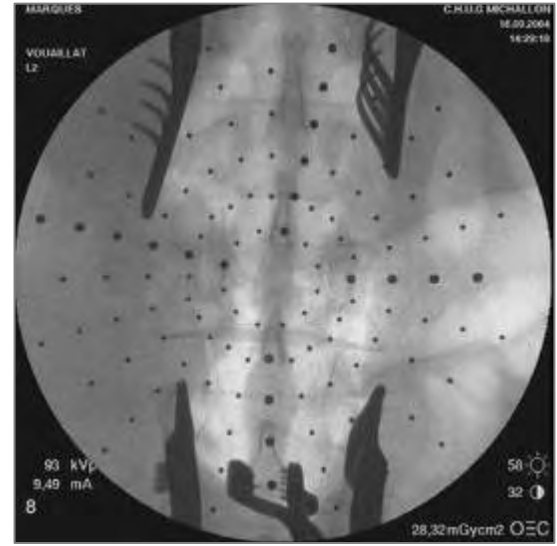

**lateral view**  **A-P view**

Fig. 4 : Two single X-ray views from A-P and lateral positions are acquired within the area who has to be operated on (here a lumbar vertebra).

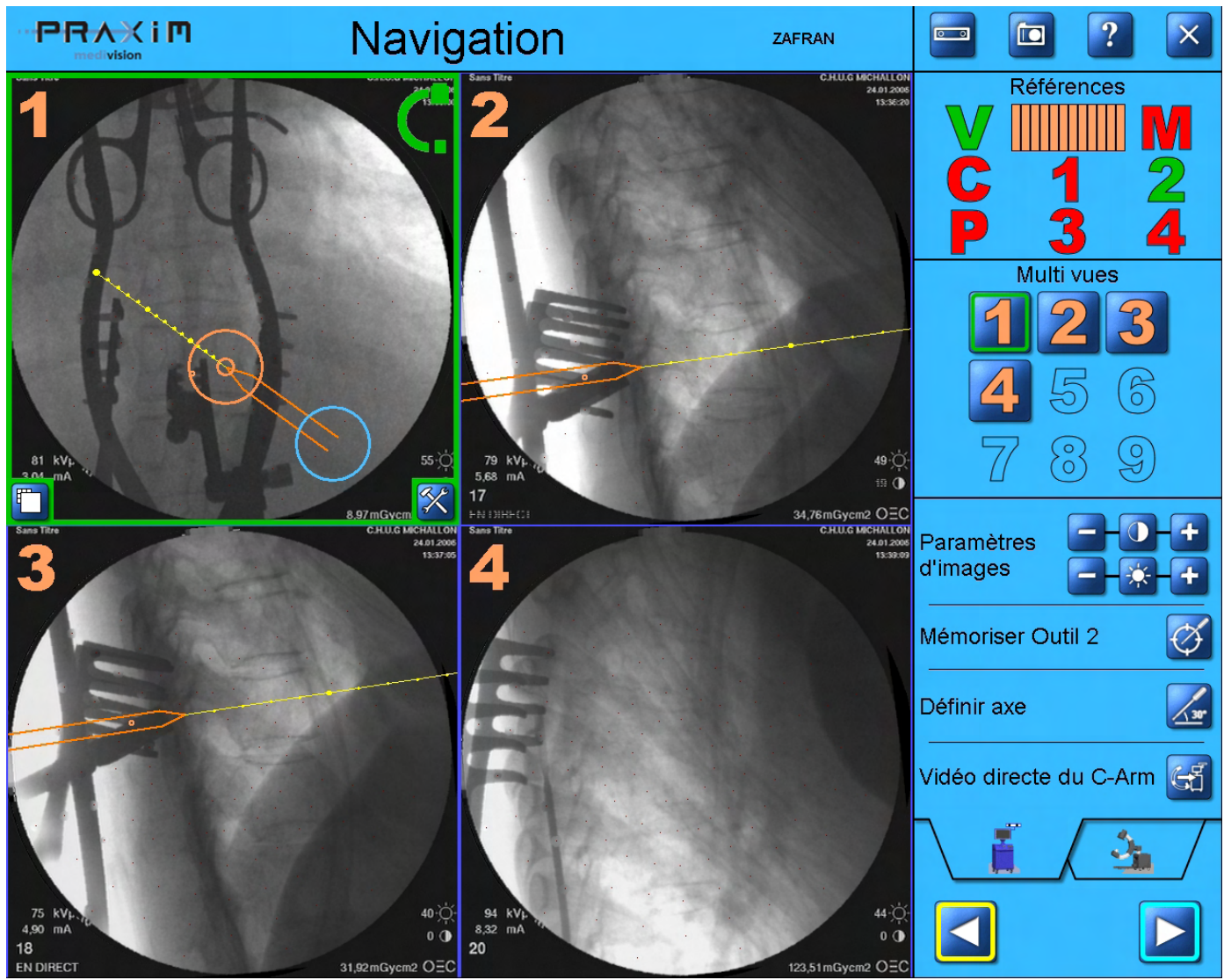

Fig. 5 : A computer-generated projection of the surgeon's surgical tool is displayed in each view. A real-time navigation in several views is possible simultaneously.

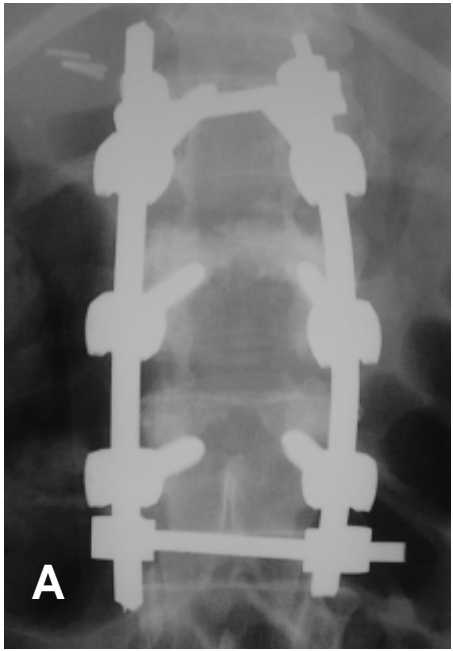
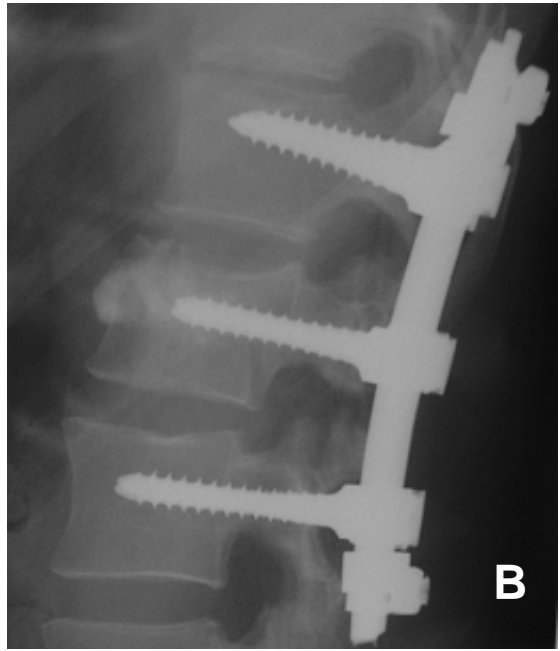
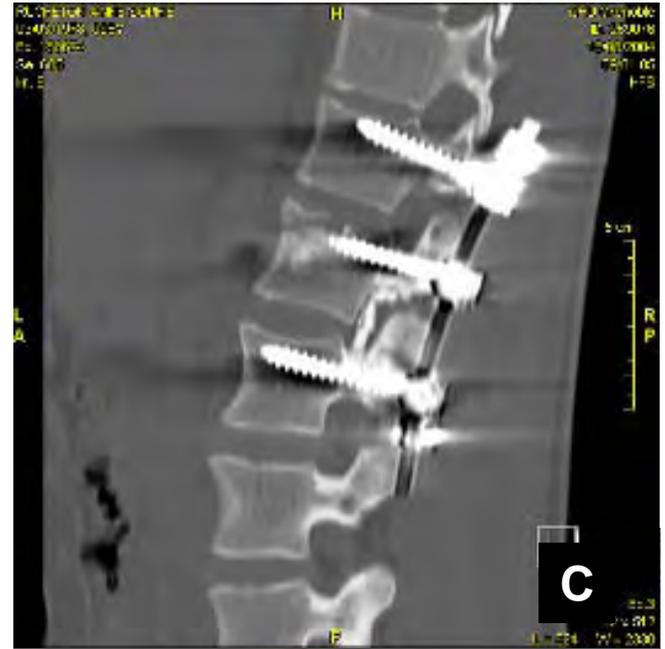
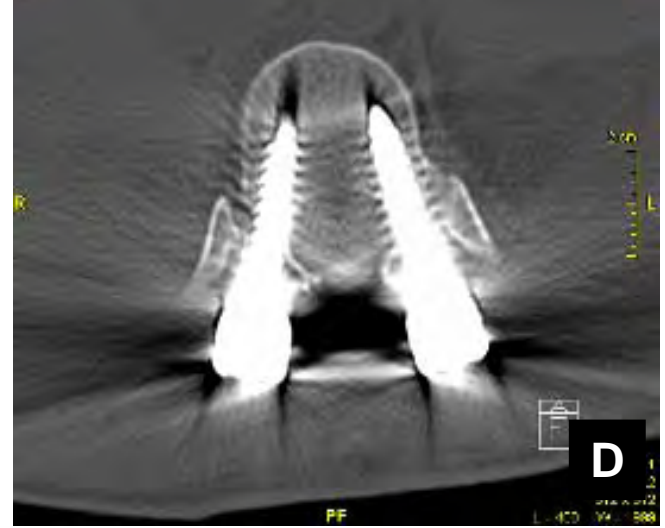

Fig. 6 : Evaluation of pedicle screw placement : L1 lumbar fracture (screws on T12, L1 and L2) (A : conventionnal AP view ; B : conventionnal lateral view)
(C : CT scan lateral view ; D : CT scan axial view on T12). The two screws on T12 are perfectly inserted in the pedicle.